\documentclass[aps,showpacs,amsmath,amssymb,pra,superscriptaddress,twocolumn,floatfix]{revtex4-1}

\usepackage{graphicx}
\usepackage{amsmath}
\usepackage{hyperref}
\usepackage{dcolumn}
\usepackage{bm}
\usepackage{color}

\usepackage[utf8]{inputenc}

\allowdisplaybreaks[1]

\begin{document}

\title{Systematic pathway to ${\mathcal{PT}}$ symmetry breaking in scattering systems}

\author{P.A. Kalozoumis}
\affiliation{Zentrum f\"ur Optische Quantentechnologien, Universit\"{a}t Hamburg, Luruper Chaussee 149, 22761 Hamburg, Germany}
\affiliation{Department of Physics, University of Athens, GR-15771 Athens, Greece}

\author{G. Pappas}
\affiliation{Department of Physics, University of Athens, GR-15771 Athens, Greece}

\author{F.K. Diakonos}
\affiliation{Department of Physics, University of Athens, GR-15771 Athens, Greece}

\author{P. Schmelcher}
\affiliation{Zentrum f\"ur Optische Quantentechnologien, Universit\"{a}t Hamburg, Luruper Chaussee 149, 22761 Hamburg, Germany}
\affiliation{The Hamburg Centre for Ultrafast Imaging, Universit\"{a}t Hamburg, Luruper Chaussee 149, 22761 Hamburg, Germany}

\date{\today}

\begin{abstract}

Recently [Phys. Rev. Lett.  {\bf 106}, 093902 (2011)] it has been shown that $\mathcal{PT}$-symmetric scattering systems with balanced gain and loss, undergo a transition from $\mathcal{PT}$-symmetric scattering eigenstates, which are norm preserving, to symmetry broken pairs of eigenstates exhibiting net amplification and loss. In the present work we derive the existence of an invariant non-local current which can be directly associated with the observed transition playing the role of an ``order parameter''. The use of this current for the description of the $\mathcal{PT}$-symmetry breaking allows the extension of the known phase diagram to higher dimensions incorporating scattering states which are not eigenstates of the scattering matrix.
\end{abstract}


\maketitle

\section{Introduction}

Quantum mechanical systems with a non-Hermitian Hamiltonian exhibiting $\mathcal{PT}$-symmetry have been the subject of intensive investigations since the work of Bender {\it et al} \cite{Bender1998,Bender2002} demonstrating that such systems can have real eigenvalues. Spontaneously breaking  the $\mathcal{PT}$-symmetry one observes a transition from a real to a complex eigenvalue spectrum \cite{Bender1998,Makris2008}. Recently it has been impressively demonstrated that $\mathcal{PT}$-symmetry can be realized in wave optical devices \cite{Makris2008, Musslimani2008a, Musslimani2008b, Guo2009, Ruter2010,Peng2014,Peng2013}. The first studies in this direction were motivated by the fact that the time-dependent Schr\"odinger equation  maps on the paraxial approximation of the electromagnetic wave equation, describing the transverse variation of the electric field~\cite{Makris2008,Ganainy2007}, where the variation on the $z$ axis plays the role of time in the corresponding Schr\"odinger equation.  Despite the fact that quantum electrodynamics is $\mathcal{T}$-invariant the classical electromagnetic theory in a medium possessing gain and/or loss leads formally to the breaking of time reversal symmetry. In optical devices with balanced gain and loss regions, located symmetrically with the respect to some mirror axis, $\mathcal{PT}$-symmetry is recovered.

In the case of electromagnetic wave propagation in optical waveguides, the $\mathcal{PT}$-breaking transition maps to that of $1$- or $2$-D bounded Schr\"odinger problems in the transverse direction. However, recently the study of light scattering in unbounded domains, where a $\mathcal{PT}$-symmetric device resides, has been addressed~\cite{Chong2011,Ge2012} and followed by an investigation of the link between the breaking of $\mathcal{PT}$-symmetry in bounded and unbounded systems~\cite{Ambichl2013}. One-dimensional $\mathcal{PT}$-symmetric  photonic heterostructures have been associated with appealing phenomena such as the existence of anisotropic transmission resonances~\cite{Ge2012}, double refraction~\cite{Makris2008} and power oscillations~\cite{Ruter2010,Zheng2010}.  Of special interest for a $\mathcal{PT}$-symmetric scatterer are the CPA-laser points~\cite{Longhi2009}, where it can act simultaneously as a coherent perfect absorber (CPA)~\cite{Chong2010} and as a laser at threshold.
An interesting development in the context of $\mathcal{PT}$-symmetric optical devices has been recently achieved \cite{Chong2011} showing that a general $\mathcal{PT}$-symmetric scattering system can undergo a multitude of spontaneous symmetry breaking transitions. A benchmark of these transitions is the parametric change of the magnitude $\vert \lambda_S \vert$ of the eigenvalues of the S-matrix leading to a phase diagram separating the $\mathcal{PT}$-symmetric, norm preserving eigenstates characterized by a pair of complex eigenvalues with $\vert \lambda_S \vert =1$ from the broken, amplified and lossy eigenstates of an inverse-conjugate pair of eigenvalues with $\vert \lambda_S \vert \neq 1$ \cite{Ge2012}. Nevertheless, the scattering states which are eigenstates of the S-matrix comprise only a small subspace of all possible scattering states, demanding specific efforts for the preparation of the appropriate conditions of the scattering system.  The question which  naturally arises then is whether there exist states which are eigenstates of the $\mathcal{PT}$-operator while they are not eigenstates of the S-matrix. Alternatively, can the phase diagram found in~\cite{Chong2011} be generalized to include scattering states which are not eigenstates of the S-matrix and if so, what would the required toolbox be such that their identification is feasible?  

In a recent work~\cite{Kalozoumis2014}, by employing a generic wave-mechanical framework, we have demonstrated how discrete symmetries such as \textit{parity} and \textit{translation} lead to symmetry-induced, non-local currents which are invariant within the spatial domains where the corresponding symmetry is obeyed (local symmetry).
These invariants, in turn, led to a mapping which was shown to
generalize the parity and Bloch theorems in the case where these symmetries are globally broken, while they are recovered in the limit of a corresponding global symmetry.       

In the present work we focus on electromagnetic wave scattering in one-dimensional $\mathcal{PT}$-symmetric optical devices and we show the existence of a non-local, spatially invariant current which can be used to classify general scattering states being eigenstates or not of the $\mathcal{PT}$-operator. In this sense our treatment generalizes the study in \cite{Chong2011} for scattering states which are not eigenstates of the S-matrix. As a result the phase diagram takes a complete and comprehensive shape, while a more direct, fundamental link to the breaking of the $\mathcal{PT}$-symmetry is achieved. The derived non-local current can be interpreted as an ``order parameter'' of the $\mathcal{PT}$-symmetry breaking transition: when it vanishes the corresponding scattering state is an eigenstate of the $\mathcal{PT}$-operator and it is non zero for states which are not $\mathcal{PT}$-eigenstates. We show that the extended phase diagram contains regions of the parameter space with scattering states which are not eigenstates of the S-matrix but norm preserving $\mathcal{PT}$-eigenstates. This has to be contrasted with the fact that S-matrix eigenstates are in general not eigenstates of the $\mathcal{PT}-$symmetry operator.

\section{ Derivation of the invariant current}
\noindent 
We consider the scattering of a monochromatic plane light wave of frequency $\omega$ from a $\mathcal{PT}$-symmetric optical multilayer in one dimension. Without loss of generality we assume that the device is elongated along the $x$-direction. The electric component of the field transverse to the $x$-axis (propagation axis) obeys the Helmholtz equation \cite{Joannopoulos2007}:
\begin{equation}
E_{xx}(x,k) +   k^2 n^2(x) E(x,k) = 0,
\label{eq:elfi}
\end{equation}
where $E_{xx}(x,k)=\frac{\partial^2 E(x,k)}{\partial x^2}$, $k=\frac{\omega}{c}$ and $n(x)$ is the spatially dependent refractive index. In the considered device $n(x)$ is in general complex indicating the gain or loss in the corresponding layer. For $\mathcal{PT}$-symmetry to hold the gain and losses have to be balanced so that $n(x)=n^*(2 a-x)$ when the center of the device is located at $x=a$. Since the scattering problem is unbounded Eq.~(\ref{eq:elfi}) has solutions for all $k \in \mathbb{R}$, although $n(x)$ is complex. Employing the method developed in \cite{Kalozoumis2014} for discrete local symmetries one can derive a spatially invariant non-local current $Q$ related to the $\mathcal{PT}$-symmetry of the refraction index $n(x)$. When the latter is globally symmetric, i.e. $n(x)=n^*(2 a-x)$ for every $x \in \mathbb{R}$, a non-vanishing $Q$ is interpreted as a remnant of the $\mathcal{PT}$-symmetry broken by the asymptotic conditions, typical for a scattering problem, as explained in \cite{Kalozoumis2014}. Particularly, to obtain $Q$ for a $\mathcal{PT}$-symmetric setup, one multiplies Eq.~(\ref{eq:elfi})  with $E^*(2 a - x,k)$ and its $\mathcal{PT}$-transform with $E(x,k)$. By subtracting, the refraction index terms cancel out and the resulting expression is written as a total derivative. This, in turn, leads to the symmetry induced, spatially invariant, non-local current:
\begin{equation}
Q=\frac{1}{2 i} \left[E^*(2 a - x, k) E_x(x,k) + E(x,k) E_x^*(2 a - x,k) \right].
\label{eq:Q}
\end{equation}
It is straightforward to show that when $Q=0$ then $E(x,k)$ is an eigenstate of the $\mathcal{PT}$-operator:
\begin{equation}
Q=0 \Rightarrow \frac{E_x(x,k)}{E(x,k)}=-\frac{\displaystyle E_x^*(2 a - x,k)}{\displaystyle E^*(2 a - x,k)}.
\label{eq:eigen1}
\end{equation}
Integrating the second equation in Eq.~(\ref{eq:eigen1}) we find:
\begin{equation}
E(x,k)=c E^*(2 a -x,k)=c \mathcal{PT}E(x,k) \Rightarrow c=\pm 1
\label{eq:eigen2}
\end{equation}
where the last equation results from the fact that $\mathcal{PT}$ is an involution ($(\mathcal{PT})^2 = I$). Thus $E(x,k)$ is an eigenstate of the $\mathcal{PT}$-operator when $Q=0$. By inserting $\pm E(x,k)$ for $E^*(2 a -x,k)$ in Eq.~(\ref{eq:Q}) one can show also the inverse, i.e. when $E(x,k)$ is a $\mathcal{PT}$-eigenstate then $Q=0$.~\footnote{Note that the internal derivative in the second term on the rhs of Eq.~(\ref{eq:Q}) introduces a relative minus sign.} Consequently $Q \neq 0$ implies that $E(x,k)$ in Eq.~(\ref{eq:Q}) is not an eigenstate of $\mathcal{PT}$ and vice versa. In a general one-dimensional scattering set-up (see Fig.~1 (a)) $Q$ can be calculated using the asymptotic expressions for the $E(x,k)$ field:
\begin{eqnarray}
E_L(x,k)&=&A e^{-ikx_L} + B e^{ikx_L}~~~;~~~x < a - \frac{L}{2} \nonumber \\
E_R(x,k)&=&C e^{-ikx_R} + D e^{ikx_R}~~~;~~~x > a + \frac{L}{2}
\label{eq:asymp}
\end{eqnarray}
with $x_L=x - a + \frac{L}{2}$ and $x_R=x - a - \frac{L}{2}$, assuming that the length of the scattering device is $L$. 

\begin{figure}[t!]
\includegraphics[width=.98\columnwidth]{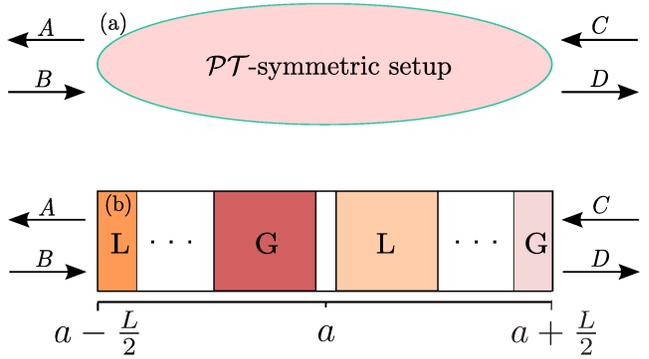}
\vspace{.2cm}
\caption{\label{fig1} (color online) (a) Representation of a generic $\mathcal{PT}-$symmetric setup. (b) General case of a $\mathcal{PT}-$symmetric photonic multilayer with an arbitrary number of slabs, exhibiting balanced gain and loss.}
\end{figure}

Inserting the asymptotic expressions (\ref{eq:asymp}) in Eq.~(\ref{eq:Q}) we obtain $Q$ in terms of the incoming and outgoing wave amplitudes as follows:
\begin{equation}
Q=k(C^* B - A D^*)
\label{eq:Qas}
\end{equation} 
For a scattering state to be a $\mathcal{PT}$ eigenstate the following condition 
--concerning its asymptotic behaviour-- should hold:
\begin{equation}
E_L^*(2 a - x,k)=\pm E_R(x,k)
\label{eq:ascon}
\end{equation}
leading to $A^*=\pm C$ and $B^*=\pm D$. These relations when inserted into Eq.~(\ref{eq:Qas}) give $Q=0$, as expected from the above discussion.

\section{Expressing $Q$ through S-matrix elements}
\noindent
In order to relate $Q$ with the parameters characterizing $n(x)$ it is useful to express the outgoing wave amplitudes $A$, $D$ in terms of the incoming amplitudes $B$, $C$ using the scattering matrix. For a general one-dimensional scattering device possessing $\mathcal{PT}$-symmetry the scattering matrix can be written as:
\begin{equation}
S=\frac{1}{\alpha} \left( \begin{array}{cc} i \beta & 1 \\ 1 & i \gamma \end{array} \right)
\label{eq:smat}
\end{equation}
with $\alpha \in \mathbb{C}$, $\beta,\gamma \in \mathbb{R}$ fulfilling the constraint $\vert \alpha \vert^2 = 1 + \beta \gamma$ \cite{Ge2012}. Using the relation defining the action of the S-matrix:
\begin{equation}
\left(\begin{array}{c} A \\ D \end{array} \right) = S \left(\begin{array}{c} B \\ C \end{array} \right)
\label{eq:inout}
\end{equation}
and Eq.~(\ref{eq:smat}) we can write the non-local invariant current $Q$ as:
\begin{equation}
Q=\frac{i k \vert C \vert^2}{\vert \alpha \vert^2} \left[2 \operatorname{Im}(\nu) - \beta
\vert \nu \vert^2  + \gamma \right],
\label{eq:QBC}
\end{equation}
where we have set $\nu=B/C$. From Eq.~(\ref{eq:QBC}) one can clearly see that $Q$ depends not only on the parameters $\alpha$, $\beta$ and $\gamma$ of the scattering matrix but also on the input amplitudes $B$, $C$. For scattering states which are eigenstates of the $\mathcal{PT}$-operator $Q$ must vanish leading to the condition:
\begin{equation}
\gamma=-2 \vert \nu \vert \sin\varphi + \beta \vert \nu \vert^2 
\label{eq:Qzero}
\end{equation}
with $\nu=\vert \nu \vert e^{i \varphi}$. Eq.~(\ref{eq:Qzero}) is a central result of the present Letter. For a general $\mathcal{PT}$-symmetric scattering set-up it determines the parameters $\alpha$, $\beta$, $\gamma$ such that the input amplitudes $B$, $C$ describe a scattering state which is a $\mathcal{PT}$ eigenstate. Thus, it can be used to classify {\em arbitrary} scattering states into eigenstates (Eq.~(\ref{eq:Qzero}) fulfilled and $Q=0$) and non-eigenstates (Eq.~(\ref{eq:Qzero}) not fulfilled and $Q \neq 0$) of the $\mathcal{PT}$-operator. 

\subsection{Relation of $Q$ to S-matrix eigenstates}
\noindent
In \cite{Chong2011} the spontaneous breaking of $\mathcal{PT}$-symmetry in photonic devices has been explored using a particular class of scattering states i.e. the eigenstates of the S-matrix. Let us note $A_e$, $B_e$, $C_e$ and $D_e$ the coefficients characterizing the asymptotic form of an eigenstate of the S-matrix in a similar way as Eq.~(\ref{eq:asymp}) does for an arbitrary scattering state. Then, since:
$$\left(\begin{array}{c} A_e \\ D_e \end{array} \right) = \lambda_S  \left(\begin{array}{c} B_e \\ C_e \end{array} \right)$$
with $\lambda_S$ being the corresponding S-matrix eigenvalue, we can rewrite Eq.~(\ref{eq:QBC}) as:
\begin{equation}
Q_e=k C_e^* B_e (1 - \vert \lambda_S \vert^2)
\label{eq:Qlam}
\end{equation}
illustrating transparently the relation between unimodularity ($\vert \lambda_S \vert =1$) of the S-matrix eigenvalues and the $\mathcal{PT}$-symmetry ($Q_e=0$) of the corresponding eigenstates. The eigenvalues as well as the magnitude of the ratio $\frac{B_e}{C_e}$ for the corresponding eigenstates of the S-matrix of a $\mathcal{PT}$-symmetric scattering system are given in terms of the parameters $\alpha$, $\beta$ and $\gamma$ in \cite{Ge2012}. Based on Eq.~(\ref{eq:Qlam}) one can determine the quantity $F_e$ which depends only on the parameters of the S-matrix:
\begin{equation}
F_e = \frac{\vert Q_e \vert}{k \vert C_e \vert^2}=\vert \nu_e \vert(1 - \vert \lambda_S \vert^2)
\label{eq:Fe}
\end{equation}
where $\nu_{e}=B_{e}/C_{e}$. This expression  becomes zero for S-matrix eigenstates which are also eigenstates of the $\mathcal{PT}$-operator, carrying the information of the symmetry breaking in the same manner as $Q_e$ does. 

\section{Phase diagram for $\mathcal{PT}$-breaking transitions in scattering}
\noindent
As discussed in \cite{Chong2011} a $\mathcal{PT}$-breaking transition occurs by tuning the parameters of the S-matrix. It has been argued that the signature of this transition is the change from unimodular to non-unimodular S-matrix eigenvalues. However as we have shown above, the conserved non-local current $Q$ is directly related to the observed transition in a transparent way, being zero for the symmetric and non-zero for the broken phase. In this sense $Q$ clearly resembles the natural ``order parameter'' of the $\mathcal{PT}$-breaking transition in a scattering system having in addition the advantage of being defined even for scattering states which are not eigenstates of the corresponding S-matrix. Thus the current $Q$ can be used to extend the existing phase diagram of the $\mathcal{PT}$-breaking transitions to arbitrary scattering states. For reasons of consistency with the literature \cite{Chong2011} we define here as the relevant ``order parameter'' the quantity $F$:
\begin{equation}
F=\frac{\vert Q \vert}{k \vert C \vert^2}
\label{eq:F}
\end{equation}
which is actually the generalization of $F_e$ to arbitrary scattering states. Using Eq.~(\ref{eq:QBC}) we can write:
\begin{equation}
F=\frac{\vert 2 |\nu| \sin\varphi - \beta |\nu|^2 + \gamma \vert}
{1 + \beta \gamma}.
\label{eq:Fabc}
\end{equation}
The condition $F=0$ determines the unbroken phase in an arbitrary $\mathcal{PT}$-symmetric scattering system while $F \neq 0$ corresponds to the broken phase. In general the phase diagram depends on four parameters $\beta$, $\gamma$, $\nu$ and $\varphi$. In the $\mathcal{PT}$-symmetric phase only three of these parameters are independent and the phase diagram will contain a connected $3D$ region (symmetric phase) characterized by $F=0$. This is clearly a richer phase diagram than that obtained by using exclusively eigenstates of the S-matrix where the variables $\nu$ and $\varphi$ are eliminated by projecting on the $\beta$, $\gamma$ plane.

Of course, this generalized phase diagram can be obtained using linear combinations of the S-matrix eigenstates since they cover the entire space of possible scattering states. However, to describe the unbroken phase via a linear combination 
\begin{equation}
\label{lin_comb} \vert\Psi\rangle=\mu_{1}\vert\Phi_{1}\rangle+\mu_{2}\vert\Phi_{2}\rangle,
\end{equation}
where $\vert \Phi_{i} \rangle,~~i=1,2$ are the S-matrix eigenstates, one has to impose the following conditions (and all their combinations):
\begin{equation}
\label{cond} \mu_{1}^{*}=\pm \mu_{1}~~~~;~~~~\mu_{2}^{*}=\pm \mu_{2},
\end{equation}
for the (in general complex) expansion coefficients $\mu_{i},~~i=1,2$. This information is optimally encoded in the single condition $Q=0$, demonstrating the propriety of the invariant current $Q$ (or equivalently $F$) as an ``order parameter'' of the $\mathcal{PT}$ breaking transition. Even more, from the physical point of view, in a scattering experiment with incidence on either side of the potential, it is expected that the ratio $\vert \nu \vert$, being a tunable parameter for a given setup, can be varied in an arbitrary manner, influencing the scattering outcome. On the other hand, restricting the incident states to S-matrix eigenstates, fixes the ratio $|\nu|$ to 1 for all the corresponding $\mathcal{PT}$-symmetric states. This can be shown by combining the conditions:
\begin{equation}
\label{s_mat_cond} A=s~B~~~;~~~D=s~C,
\end{equation}
which emerge from the fact that the incoming state is an eigenstate of the S-matrix ($s$ being the respective eigenvalue), with 
\begin{equation}
\label{s_mat_cond} B^{*}=\Lambda~D~~~;~~~A^{*}=\Lambda~C,
\end{equation}
which stem from the $\mathcal{PT}$-symmetry of the states. $\Lambda$ (which is a phase) is the corresponding eigenvalue.

\begin{figure}[t!]
\includegraphics[width=.99\columnwidth]{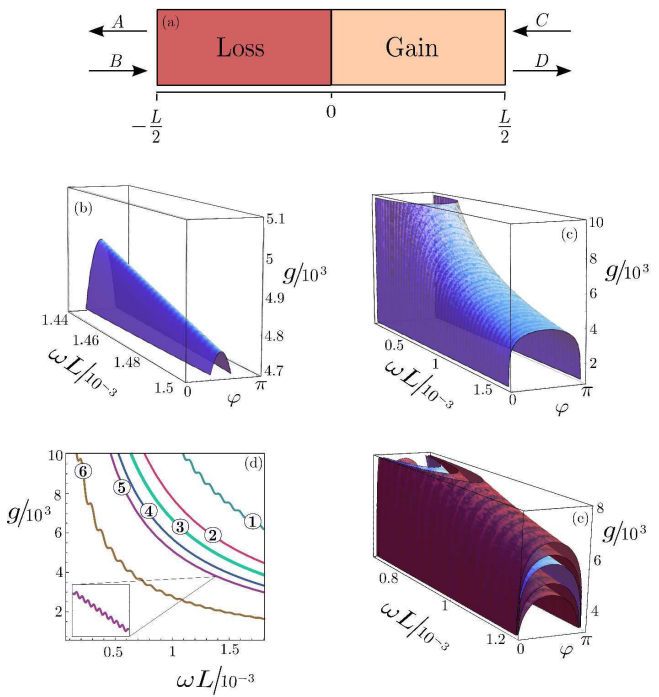}
\vspace{.2cm}
\caption{\label{fig2} (color online) (a) The considered $\mathcal{PT}$-symmetric scattering device comprised of two attached dielectric materials with balanced loss and gain, respectively. The refraction index of both slabs is $n_{0}=3$ and is equal  to the refraction index $n$ in the asymptotic regions on either side of the setup $(n=n_{0})$. (b) Phase diagram in a three-dimensional parameter space, corresponding to eigenstates of the S-matrix. (c) The same phase diagram as in (b) for a broader parameter range. (d) Limiting phase diagram lines corresponding to angle $\varphi=\pi/2$ (phase of the ratio of the incoming amplitudes). Curves $(1)$, $(2)$, $(4-6)$ correspond to $|\nu|=0.01,~0.3,~3,~6,~100$ (non-S-matrix eigenstates), whereas curve  $(3)$ is characterized by $|\nu_{3}|=1$ (S-matrix eigenstates). (e) Three-dimensional phase diagrams, emerging from Eq.~(\ref{eq:Qzero}), corresponding to (bottom to top) $|\nu|=6,~3,~1,~0.3,~0.6$. The  blue surface with $|\nu|=1$ (middle) corresponds to the case of the S-matrix eigenstates  and is part of the surface shown in (c).}  
\end{figure}

To illustrate the above issues in a more transparent way we construct the phase diagram for the specific $\mathcal{PT}$-symmetric scattering problem considered in Refs.~\cite{Chong2011,Ge2012,Ambichl2013}. The device, shown in Fig.~\ref{fig2} (a) is comprised of two attached dielectric materials, one inducing losses ($n_{0}+ig$) and the other gain ($n_{0} - ig$). The value of the refraction index $n_{0}$ coincides with the respective value $n$ in the asymptotic regions on either side of the device ($n_{0}=n$).  Figures~\ref{fig2} (b), (c) illustrate the phase diagram of the corresponding setup for incoming waves which are all eigenstates of the S-matrix. Particularly, subfigure (b) shows, for reasons of comparison, the phase diagram for a limited parameter range, similar to that used in Fig.~2 (a) of Ref.~\cite{Chong2011}, while (c) illustrates the same situation for a broader parameter range.  The emerging surfaces stem from the condition given in Eq.~(\ref{eq:Qzero}) and the corresponding parameters which determine its form are the gain/loss rate ($g$), the scaled frequency $\omega L$ and the angle $\varphi$, appearing in the phase of $\nu$. Note, that $g$ and $\omega L$ appear in the expressions of $\gamma$ and $\beta$, contained in the S-matrix. This $3D$ representation is advantageous in several ways, revealing transparently new aspects of the structure of the phase diagram. In the first place, there are no solutions of Eq.~(\ref{eq:Qzero}) for $\pi<\varphi<2\pi$ and consequently in this regime of angles $\mathcal{PT}$-symmetric states do not exist. Additionally, the  $3D$ phase diagram is symmetric around $\varphi=\frac{\pi}{2}$, the latter being the limiting line in the $2D$ $g-\omega L$ plane, in the sense that the lines which correspond to the phase angles $0<\varphi<\frac{\pi}{2}$, equally project in the range $\frac{\pi}{2}<\varphi<\pi$. Therefore, projecting the diagrams (b) and (c) onto the $g-\omega L$ plane, the phase angle $\varphi=\pi/2$ forms a limiting curve separating $\mathcal{PT}$-symmetric (below this curve) from  $\mathcal{PT}$ non-symmetric (above this curve) S-matrix eigenstates. With this approach we recover the phase diagram derived in Ref.~\cite{Chong2011}. However, since  every $\mathcal{PT}$-symmetric state, which is also an eigenstate of the S-matrix, has $|\nu|=1$, the phase $\varphi$ of $\nu$ provides essential information concerning the loci where $\mathcal{PT}$-symmetry is obeyed. This information will have particular impact on the preparation of a scattering system in the $\mathcal{PT}$-symmetric phase.  In Fig.~\ref{fig2} (d) we show the limiting lines which correspond to $\varphi=\pi/2$, for six different values of $|\nu|$. Curve ($3$) corresponds to incoming states which are eigenstates of the S-matrix ($|\nu|=1)$. The existence of $\mathcal{PT}$-symmetric states \textit{which are not eigenstates} of the S-matrix is obvious.  As we can see, the parameters $\omega L, g$ cannot uniquely describe the properties of a generic scattering state with respect to $\mathcal{PT}$-symmetry, since the limiting curve depends also on the ratio of the incoming amplitudes $\nu$.
The limiting curves ($1,~2$) and ($4-6$), are characterized by $|\nu| \neq 1$,  describing generic $\mathcal{PT}$-symmetric states which are not S-matrix
eigenstates. The oscillatory structure, observed in all curves (inset), is attributed to the sinusoidal terms which involve the phase diagram parameters and appear in the form of the S-matrix elements $\beta$ and $\gamma$. Finally, in Fig.~\ref{fig2} (e), the full, $3D$ phase diagrams are shown for several values of $|\nu|$. The blue surface corresponds to the S-matrix eigenstates case and is part of the surface shown in (c). Obviously, the $\mathcal{PT}$-symmetric S-matrix eigenstates cover only a small fraction of all possible $\mathcal{PT}$-symmetric states.

\section{Concluding remarks}
\noindent
The central theme in the present work has been the derivation of a spatially invariant non-local current $Q$ for the Helmholtz equation with $\mathcal{PT}$-symmetry. This current offers a natural ``order parameter'' for the spontaneous $\mathcal{PT}$-symmetry breaking transitions in one-dimensional scattering. There are two striking features of $Q$: (i)  it provides a direct link to the violation ($Q \neq 0$) or not ($Q=0$) of the global $\mathcal{PT}$-symmetry and (ii) it allows the study of the $\mathcal{PT}$-symmetry breaking transitions with scattering states which are not eigenstates of the S-matrix, enriching significantly the associated phase diagram. This extension, in turn, allows for the observation and manipulation of physical properties associated to the $\mathcal{PT}$-symmetry of the setup (anisotropic transmission resonances, CPA-laser solutions), in parametric regions, which were previously inaccessible.
Last but not least, it must be noticed that $Q$ exists also for setups where the $\mathcal{PT}$-symmetry holds only within a finite spatial domain (as a spatially invariant quantity within this domain). This opens the intriguing perspective to study  novel transitions in systems where the $\mathcal{PT}$-symmetry is broken globally but retained locally in well-defined spatial regions, leading possibly to an extended control of the phase diagram. Additionally, since the identification of local symmetries in globally non-symmetric, aperiodic systems has led to setups with prescribed perfect transmission resonance properties~\cite{Kalozoumis2013a} and their fundamental classification~\cite{Kalozoumis2013}, one could expect an extended variability of anisotropic transmission resonances in the case where $\mathcal{PT}$-symmetry is fulfilled only in restricted spatial domains.

We thank A.V. Zampetaki, C. Morfonios, V. Zampetakis and M. Diakonou for fruitful and illuminating discussions.


\begin{thebibliography}{99}

\bibitem{Bender1998} C. M. Bender and S. Boettcher, Phys. Rev. Lett. {\bf 80}, 5243 (1998).

\bibitem{Bender2002} C. M. Bender, D. C. Brody, and H. F. Jones, Phys. Rev. Lett. {\bf 89}, 270401 (2002).

\bibitem{Makris2008} K. G. Makris, R. El-Ganainy, D. N. Christodoulides and Z. H. Musslimani, Phys. Rev. Lett. {\bf 100}, 103904 (2008); Phys. Rev. A {\bf 81}, 063807 (2010).

\bibitem{Musslimani2008a} Z. H. Musslimani, K. G. Makris, R. El-Ganainy, and D. N. Christodoulides, Phys. Rev. Lett. {\bf 100}, 030402 (2008); 

\bibitem{Musslimani2008b} Z. H. Musslimani, K. G. Makris, R. El-Ganainy, and D. N. Christodoulides, J. Phys. A {\bf 41}, 244019 (2008).

\bibitem{Guo2009} A. Guo, G. J. Salamo, D. Duchesne, R. Morandotti, M. Volatier-Ravat, V. Aimez, G. A. Siviloglou, and D. N. Christodoulides, Phys. Rev. Lett. {\bf 103}, 093902 (2009).

\bibitem{Ruter2010} Christian E. R\"uter, Konstantinos G. Makris, Ramy El-Ganainy, Demetrios N. Christodoulides, Mordechai Segev and Detlef Kip, Nat. Phys. {\bf 6}, 192 (2010).

\bibitem{Peng2014} B. Peng,	S. K. \"Ozdemir,	F. Lei,	F. Monifi,	M. Gianfreda,	G. L. Long,	S. Fan,	F. Nori,	C. M. Bender	 and L. Yang, Nat. Phys. \textbf{10}, 394 (2014).

\bibitem{Peng2013} C. M. Bender, M. Gianfreda, S. K. \"Ozdemir, B. Peng, and L. Yang,  Phys. Rev. A \textbf{88}, 062111 (2013).

\bibitem{Ganainy2007} R. El-Ganainy, K. G. Makris, D. N. Christodoulides and Z. H. Musslimani, Opt. Lett. \textbf{32}, 2632 (2007).

\bibitem{Chong2011} Y. D. Chong, L. Ge, and A. D. Stone, Phys. Rev. Lett.  {\bf 106}, 093902 (2011).

\bibitem{Ge2012} L. Ge, Y. D. Chong, and A. D. Stone, Phys. Rev. A {\bf 85}, 023802 (2012). 
 
\bibitem{Ambichl2013} P. Ambichl, K. G. Makris, L. Ge, Y. D. Chong, A. D. Stone, and S. Rotter, Phys. Rev. X {\bf 3}, 041030 (2013).

\bibitem{Zheng2010} M. C. Zheng,  D. N. Christodoulides, R. Fleischmann and T. Kottos, Phys. Rev. A \textbf{82}, 010103(R) (2010).

\bibitem{Longhi2009} S. Longhi, Phys. Rev. A \textbf{82}, 031801(R) (2010).

\bibitem{Chong2010} Y. D. Chong, Li Ge, H. Cao and A. D. Stone, Phys. Rev. Lett. \textbf{105}, 053901 (2010).

\bibitem{Kalozoumis2014} P. A. Kalozoumis, C. Morfonios, F. K. Diakonos and P. Schmelcher, Phys. Re. Lett. \textbf{113}, 050403 (2014).
 
\bibitem{Joannopoulos2007} J. D. Joannopoulos, R. D. Meade, and J. N. Winn, {\it Photonic Crystals: Molding the Flow of Light}, Princeton University Press, Princeton, NJ (2007).

\bibitem{Kalozoumis2013} P. A. Kalozoumis, C. Morfonios, N. Palaiodimopoulos, F. K. Diakonos, and P. Schmelcher, Phys. Rev. A {\bf 88}, 033857 (2013).

\bibitem{Kalozoumis2013a} P. A. Kalozoumis, C. Morfonios, F. K. Diakonos, and P. Schmelcher, Phys. Rev. A {\bf 87}, 032113 (2013).



\end{thebibliography}
\end{document}